\documentclass[twocolumn,amsmath,amssymb]{revtex4}
\usepackage{bm}
\usepackage{latexsym}
\usepackage{dcolumn}
\usepackage{amsmath,amsfonts,amssymb}
\usepackage{graphicx,epsfig}
\usepackage{color}
\usepackage{psfrag}
\usepackage{amsthm}
\def\be{\begin{equation}}
\def\ee{\end{equation}}
\def\bes{\begin{equation*}}
\def\ees{\end{equation*}}
\usepackage{slashed}
\def\met{\slashed E_T}
\begin{document}
\title{Searching for Charged Higgs Bosons in the $B-L$ Supersymmetric Standard Model \\ at the High Luminosity Large Hadron Collider}
\author{W. Abdallah$^{1,2}$, A. Hammad$^{3}$, S. Khalil$^{4}$ and S. Moretti$^{5}$}
\affiliation{
$^1$Harish-Chandra Research Institute, Chhatnag Road, Jhunsi, Allahabad 211019, India.\\
$^2$Department of Mathematics, Faculty of Science, Cairo University, Giza 12613, Egypt.\\
$^3$Department of Physics, University of Basel, Klingelbergstra\ss e 82, CH-4056 Basel, Switzerland.\\
$^4$Center for Fundamental Physics, Zewail City of Science and Technology, 6 October City, Giza 12588, Egypt.\\
$^5$School of Physics and Astronomy, University of Southampton, Highfield, Southampton SO17 1BJ, UK.}
\begin{abstract}
\noindent
Upon assuming the $B-L$ Supersymmetric Standard Model (BLSSM) as theoretical framework accommodating a multi-Higgs sector, we
assess the scope of the High Luminosity Large Hadron Collider ({{HL-LHC}}) in accessing charged Higgs bosons ($H^\pm$) produced in pairs from $Z'$ decays. We show that, by pursuing both di-jet and tau-neutrino decays, several signals can be established for $H^\pm$ masses ranging from about $M_{W}$ to above $m_t$ and $Z'$ masses between 2.5~TeV and 3.5~TeV.  The discovery can be attained, in a nearly background free environment in some cases, owing to the fact that the very massive resonating $Z'$ ejects the charged Higgs bosons at very high transverse momentum, a kinematic region where any SM noise is hugely depleted.
\end{abstract}
\maketitle
Searches for light charged Higgs bosons $(H^\pm)$ in the decay of top
quarks, $t \to H^\pm b$, are presently being carried
out at the Large Hadron Collider (LHC), with the assumption that their decay channels are dominated by $H^\pm \to \tau\nu_\tau$ or $H^\pm \to jj$, where $j$ represents a jet and the possible partonic combinations are $cs$ and $cb$. For heavy $H^\pm$ states, with $M_{H^\pm}>m_t$, one resorts instead to the $H^\pm\to tb$ channel, via associated production of a charged Higgs boson with a top quark. (See   
\cite{BergeaasKuutmann:2017yud,Laurila:2017phk} for recent reviews by ATLAS and CMS.) In both cases then, $H^\pm$ states are produced in single mode. The scope for testing charged Higgs boson pair production at the LHC is instead much limited, whichever the channel to be pursued \cite{Moretti:2001pp}, primarily owing to the small cross sections involved.  The experimental analyses are carried out model independently. The results though can  be interpreted in a variety of Beyond the SM (BSM) scenarios (see
\cite{Akeroyd:2016ymd} for a recent review).  A popular framework in this respect is the Minimal Supersymmetric Standard Model (MSSM), which is the most economical realization of Supersymmetry (SUSY) containing $H^\pm$ states. However, this SUSY incarnation is plagued by innumerable problems, both theoretical and experimental, so that non-minimal models of SUSY are being explored  \cite{Book}. 

Amongst these, an intriguing one is the $B-L$ Supersymmetric Standard Model (BLSSM), which, while inheriting the beneficial aspects of SUSY from the MSSM, it surpasses it as it naturally predicts massive neutrinos (as required by experiment), an enlarged Higgs sector (which allows for a SM limit) and an expanded  gauge symmetry (potentially a remnant of a Grand Unification Theory (GUT)) \cite{Khalil:2012gs,Book} as well as a Dark Matter (DM) candidate (the SUSY counterpart of a neutrino, i.e., a sneutrino) that, thanks to its interactions with richer Higgs and gauge spectra, complies with both direct and indirect constraints better than the MSSM candidate \cite{DelleRose:2017ukx,Khalil:2011tb,DelleRose:2017uas,Abdallah:2017gde}.     
 
\begin{figure}[h]
\begin{center}
\vspace{-0.7cm}
\includegraphics[width=0.45\textwidth]{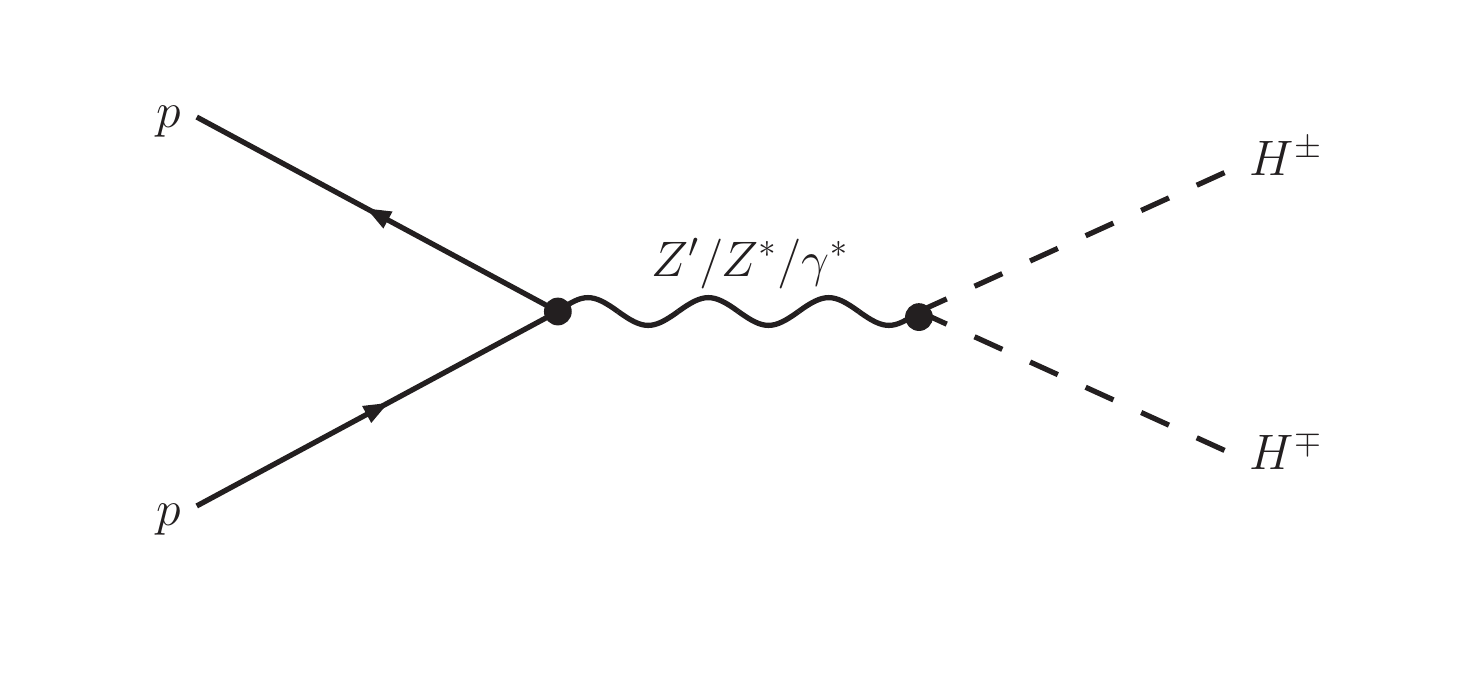}
\vspace{-0.7cm}
\caption{Feynman diagram of charged Higgs production intermediated by $Z'/Z^*/\gamma^*$.}
\vspace{-0.7cm}
\label{fig:feyn}
\end{center}
\end{figure}

\begin{figure}[h]
\vspace{-0.2cm}
\begin{center}
\includegraphics[width=0.5\textwidth]{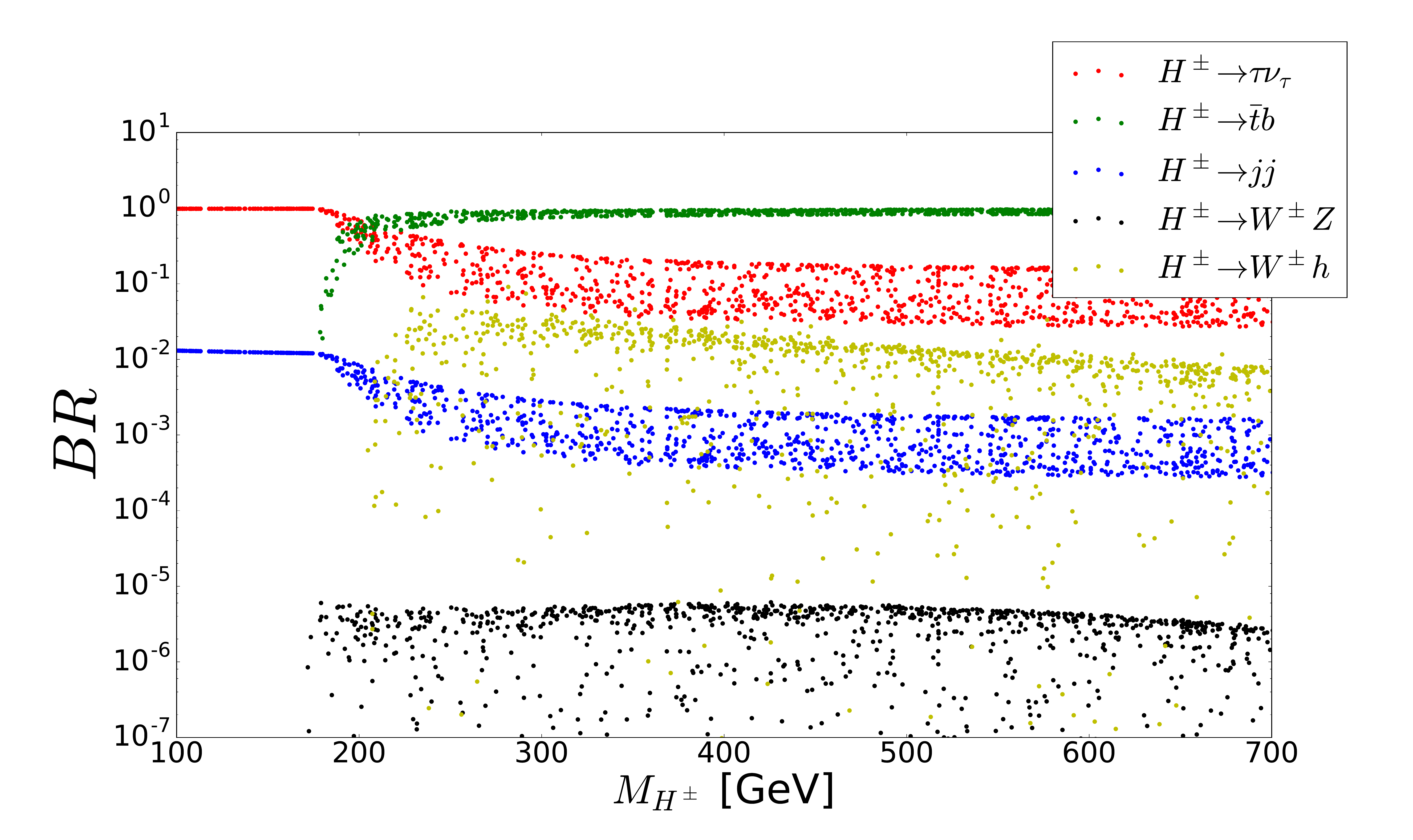}
\vspace{-0.5cm}
\caption{BRs of the charged Higgs boson in the BLSSM.}
\label{fig:br}
\vspace{-0.7cm}
\end{center}
\end{figure}

The BLSSM  is also an example of New Physics (NP) that predicts the existence of charged Higgs bosons. While single $H^\pm$ production here is not dissimilar from the MSSM case, a notable difference emerges in the case of double  $H^\pm$ production. The reason why MSSM cross sections at the LHC for $pp\to H^+H^- X$ processes are small is that charged Higgs boson pairs are never produced resonantly. This is unlike the BLSSM, where the condition $M_{Z'}>2M_{H^\pm}$ is naturally realized, given the constraints on the $Z'$ mass, around 3.5~TeV presently \cite{DelleRose:2017ukx,DelleRose:2017uas}. Hence, within the BLSSM, once can resort to the  $pp\to Z^\ast,\gamma^\ast, Z'\to H^+ H^-$ mode, see
 Fig.~\ref{fig:feyn},
wherein the $Z'$ (resonant) component, together with its interference with the SM, is the actual signal and the $\gamma^*,Z^*$ ones are the (irreducible) background ones, which are non-resonant given that the current lowest mass limit on $H^\pm$ states is essentially the $W^\pm$ mass \cite{Akeroyd:2016ymd}. Such a signal is best searched for via the aforementioned $\tau\nu_\tau$ and $jj$ channels, even when $M_{H^\pm}>m_t$, as efficiency of other decay modes is much poorer in comparison. In the light of this, there are also reducible backgrounds to be dealt with, primarily $t\bar t$, gauge boson pair production ($W^+W^-$ and $ZZ$) and $W^\pm, Z$~+~jets. The Branching Ratios (BRs) of the charged Higgs boson of the BLSSM can be seen in   Fig.~\ref{fig:br}, wherein the BLSSM points have been generated over the following intervals of its fundamental parameters:
{{$0.5\le \mu\le 3~{\rm TeV},~50\le M_A\le 10^3~{\rm TeV},~10\le \tan\beta\le 30,~0.3\le g_{BL}\le 0.75,~-0.3\le \tilde{g}\le -0.2,~M_1= 1.5~{\rm TeV},~M_2= 1.5~{\rm TeV},~M_3= 3.5~{\rm TeV},~M_{A^\prime}= 10^2~{\rm TeV},~\mu^\prime= 0.6~{\rm TeV.}$}}

At this point, it is also worth mentioning that other $H^+H^-$ production modes exist in the BLSSM that could play a role at the LHC in the context we are addressing, specifically, induced by gluon-gluon induced channels. These can be neglected though for our purposes, as the box component does not benefit from any BLSSM specific enhancement while the triangle  one (which would indeed include a $Z'$ boson in $s$-channel) is suppressed owing to the Landau-Yang mechanism \cite{Moretti:2014rka}. Further, also BLSSM intrinsic backgrounds, such as $Z^\prime\to W^+ W^-$ and $Z^\prime\to W^\pm H^\mp$ decays are negligible, as they are proportional to (the sine of) the $Z$-$Z'$ mixing angle, $\sin\theta^\prime$, which is constrained by LEP data to be less than $10^{-3}$~\cite{ALEPH:2005ab,Accomando:2016sge}. 
\begin{figure}[h]
\begin{center}
\includegraphics[width=0.235\textwidth]{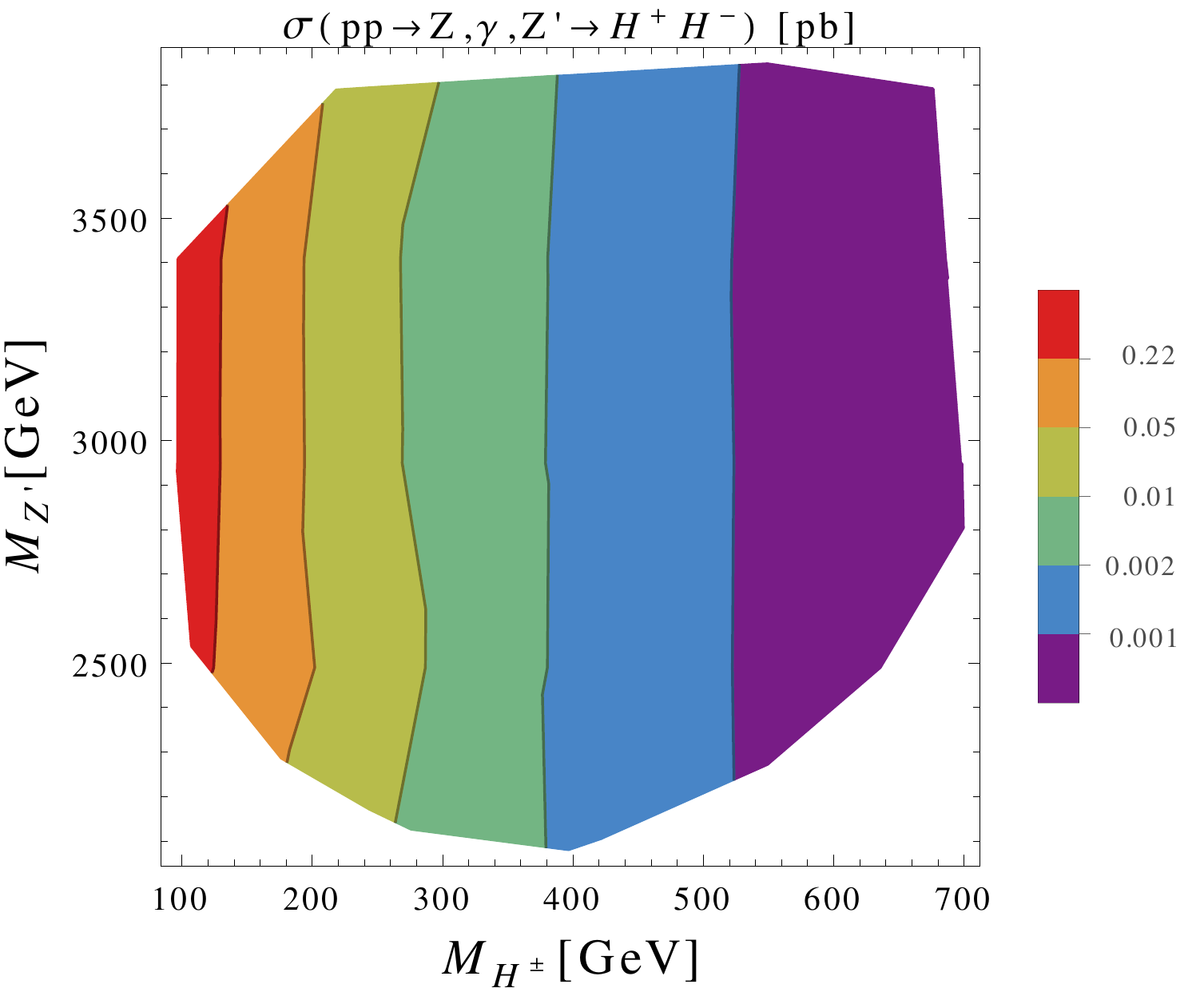}
\includegraphics[width=0.235\textwidth]{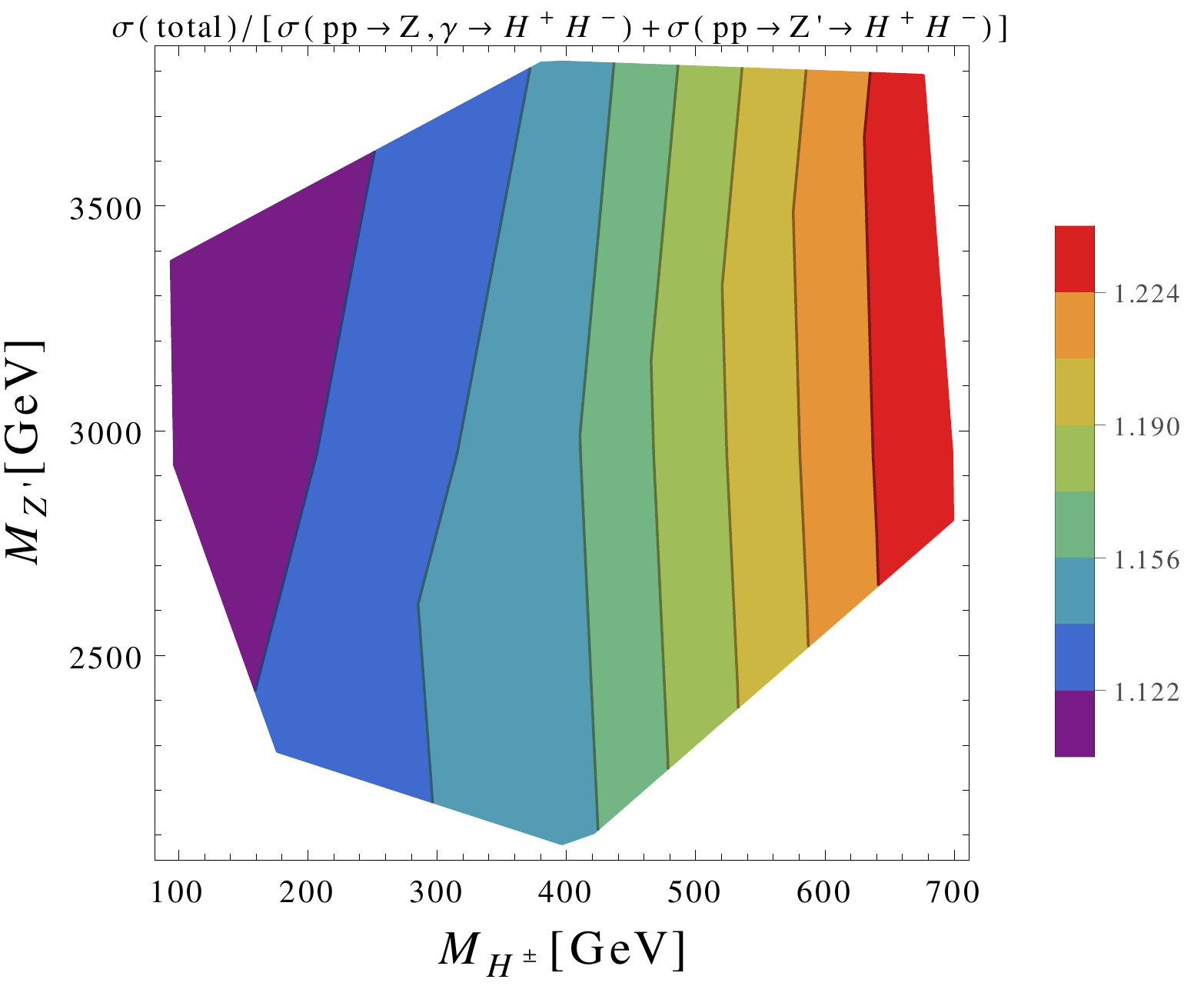}
\caption{Absolute cross section in pb  for  $\sigma(pp\to\gamma^*,Z^*,Z'\to H^+H^-)$ (left frame) and
 ratio $\sigma({pp\to\gamma^*,Z^*,Z'\to H^+H^-})/[\sigma(pp\to\gamma^*,Z^*\to H^+H^-)+\sigma(pp\to Z'\to H^+H^-)]$ (right frame) at 14~TeV, after the cut $|M_{Z'}-M_{H^+H^-}|\leq 15~\Gamma_{Z'}$.}
\label{fig:Xsection-scan}
\vspace{-0.4cm}
\end{center}
\end{figure}

\begin{figure}[t]
\begin{center}
\includegraphics[width=0.45\textwidth]{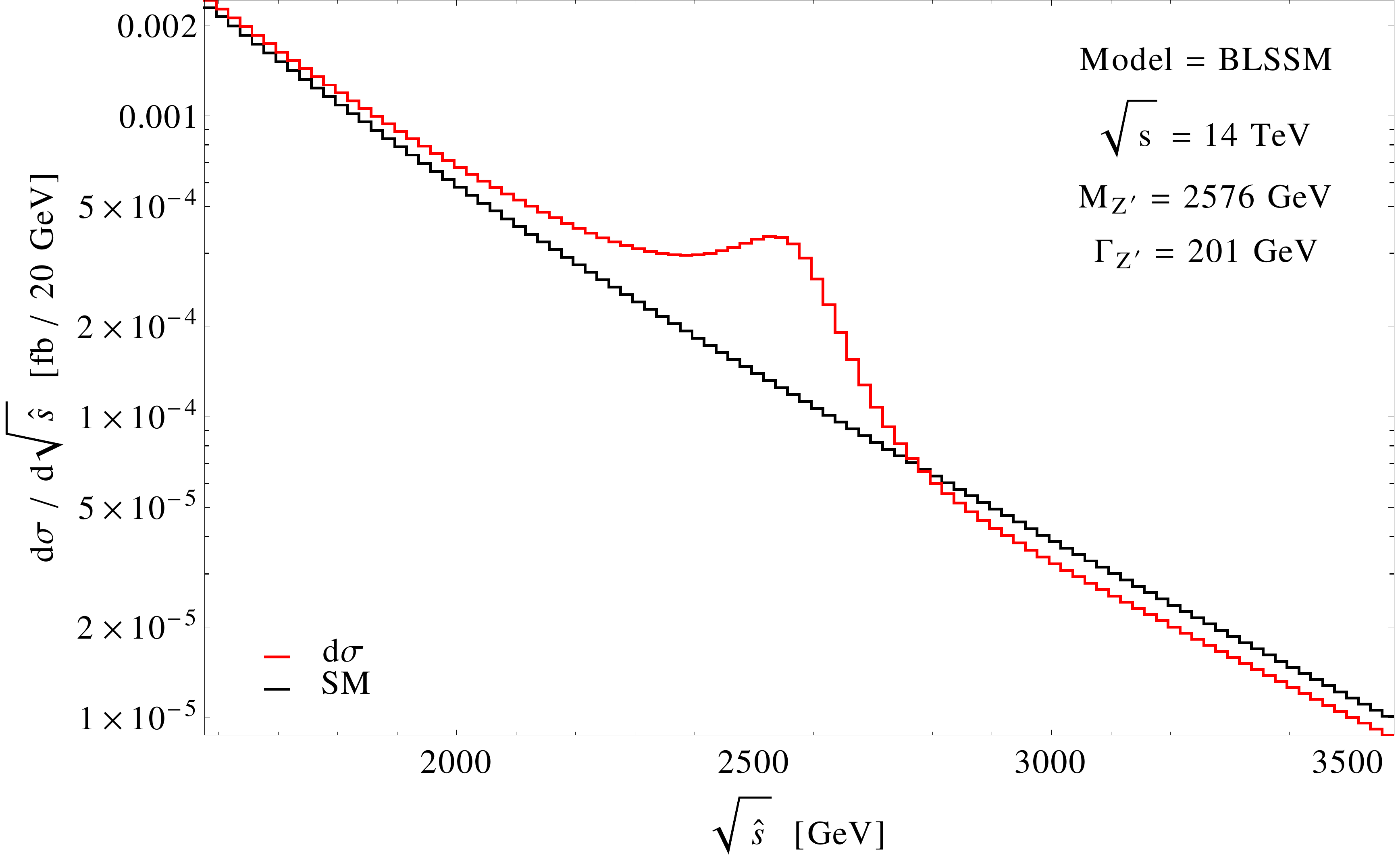}\\[0.5cm] 
\includegraphics[width=0.45\textwidth]{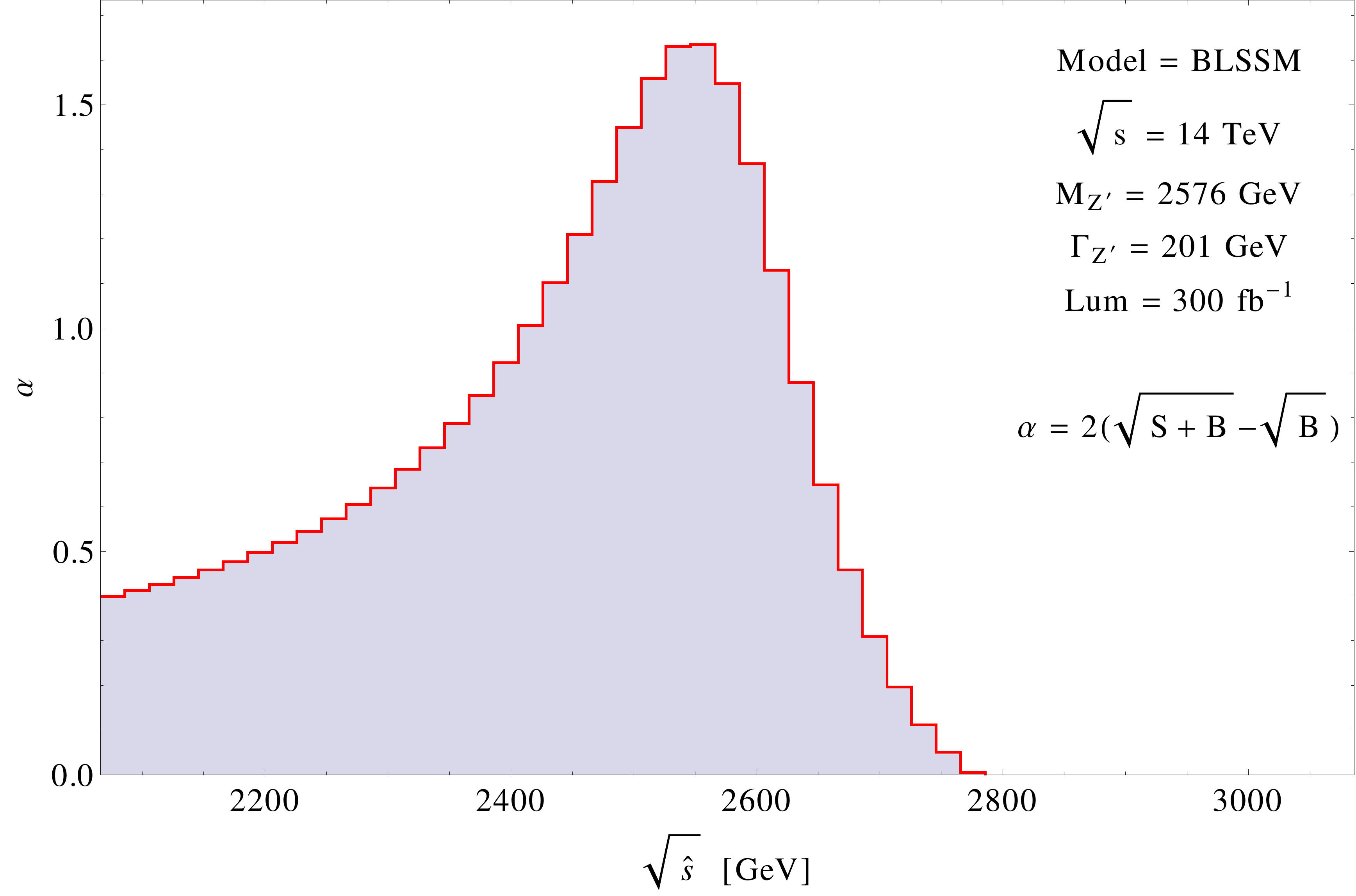}
\vspace{-0.3cm}
\caption{Top panel: Differential cross section distribution at leading order for the BLSSM process $pp\to \gamma^\ast,Z^\ast,Z^\prime\to {{l^+ l^-}}$  at 
BP3 of Tab.~\ref{table:0} (red) versus the one for the SM channel $pp\to \gamma^\ast,Z^\ast\to {{l^+ l^-}}$  (black). Bottom panel: The corresponding significance, $\alpha$, of the  {{$Z'$}} signal {{($S$)}} versus the (irreducible) background {{$pp\to \gamma^\ast,Z^\ast\to l^+ l^-$ ($B$), where $S$ is identified as the difference between the yield of the subprocess $pp\to \gamma^\ast,Z^\ast,Z^\prime\to l^+ l^-$ and $B$.}}
(See \cite{Abdallah:2015uba} for the definition of $\alpha$.)}
\label{fig:signal_strength}
\vspace{-0.7cm}
\end{center}
\end{figure}

Another aspect that renders $H^+H^-$ production at the LHC within the BLSSM more interesting than in  the MSSM  is the fact that, while in the latter the $\gamma H^+H^-$ and $Z H^+H^-$ vertices are fixed by the SM gauge symmetries, in the former one has some freedom  to find a sizeable range of parameters that can make the $Z'H^+H^-$ coupling sufficiently large to offset the phase space suppression coming with the fact that the $Z'$ is bound to be rather heavy, as discussed. In the BLSSM, the coupling of charged Higgs bosons to the  $Z'$ is generated through the possible mixing in the mass matrix of the $Z$ and $Z'$ gauge bosons and/or the kinetic mixing between $U(1)_Y$ and $U(1)_{B-L}$, which is $\sim \tilde{g}$ (see  Ref.~\cite{OLeary:2011vlq} for further details).  Over the above volume of BLSSM parameter space, we find cross sections  $\sigma(pp\to\gamma^*,Z^*,Z'\to H^+H^-)$, i.e., prior to any $H^\pm$ decay restricted to the kinematic range $|M_{Z'}-M_{H^+H^-}|\leq 15~\Gamma_{Z'}$,  given in Fig.~\ref{fig:Xsection-scan} (left frame) over the $(M_{Z'},M_{H^\pm})$ plane for the fixed value of $\tilde g=-0.29$. It is important to notice here that, in the definition of the BLSSM signal, a key role is played by the interference between the $Z'$ and $\gamma^*,Z^*$ components of the process, which turns out to be constructive over the relevant parameter range, as can be seen from Fig.~\ref{fig:Xsection-scan} (right frame), {{where we plot the ratio $\sigma(pp\to\gamma^*,Z^*,Z'\to H^+H^-)/[\sigma(pp\to\gamma^*,Z^*\to H^+H^-)+\sigma(pp\to Z'\to H^+H^-) ]$}}. From those in this plot, we now select five Benchmark Points (BPs), which differ in the $Z'$ and $H^\pm$ masses but have common gauge couplings $g_{B-L}$ and $\tilde g$, see Tab.~\ref{table:0}, to be used in the forthcoming phenomenological analysis. In defining these, we have made sure that, on the one hand, they do not fall foul of the aforementioned LEP (indirect) constraints and, on the other hand, the ensuing $Z'$ will not have been discovered via LHC (direct) searches in Drell-Yan (DY) mode already (i.e., by the time the analysis that we advocate will be pursued), which is demonstrated by  
Fig.~\ref{fig:signal_strength} for the illustrative case of BP3 (it is the same for the other BPs as well) \cite{Abdallah:2015uba}. In fact, the top frame herein shows  the line-shape of the differential cross section mapped in invariant mass of the final state $M_{ll}\equiv \sqrt{\hat s}$ ($l=e,\mu$) near the $Z'$ resonance, here of 2576~GeV,  wherein the 
$Z'$ effect and that of its interference with the irreducible SM background are clearly visible against the latter,  yet, the significance of these two contributions over the SM noise after, e.g., 300~fb$^{-1}$ of luminosity is about 1.5 at best, see the bottom frame. Even for a tenfold increase in collider luminosity, as expected at the  High Luminosity LHC (HL-LHC) \cite{Gianotti:2002xx}, corresponding to an increase of a factor $\sqrt{10}$  in significance, the latter should remain below~5.    

\begin{table}[!h]
\begin{center}
\begin{tabular}{|c|c|c|c|c|c|}
\hline
 & $M_{Z^\prime}$ [GeV]& $\Gamma_{Z^\prime}$ [GeV]&$M_{H^\pm}$ [GeV]& $g_{B-L}$ & $\tilde{g}$\\
\hline 
BP1 & 2576  &201& 201  & 0.35 & $-0.29$  \\
\hline
BP2 & 2576  &201& 175  & 0.35 & $-0.29$  \\
\hline
BP3 & 2576  &201& 143  & 0.35 & $-0.29$  \\
\hline
BP4 & 2576  &201& 94  & 0.35 & $-0.29$  \\
\hline
BP5 & 3380  &315& 141  & 0.35 & $-0.29$  \\
\hline
\end{tabular}
\caption{Benchmark Points (BPs) that we will use.} 
 \label{table:0}
\end{center}
\end{table}

\begin{figure}[t]
\begin{center}
\includegraphics[width=0.52\textwidth]{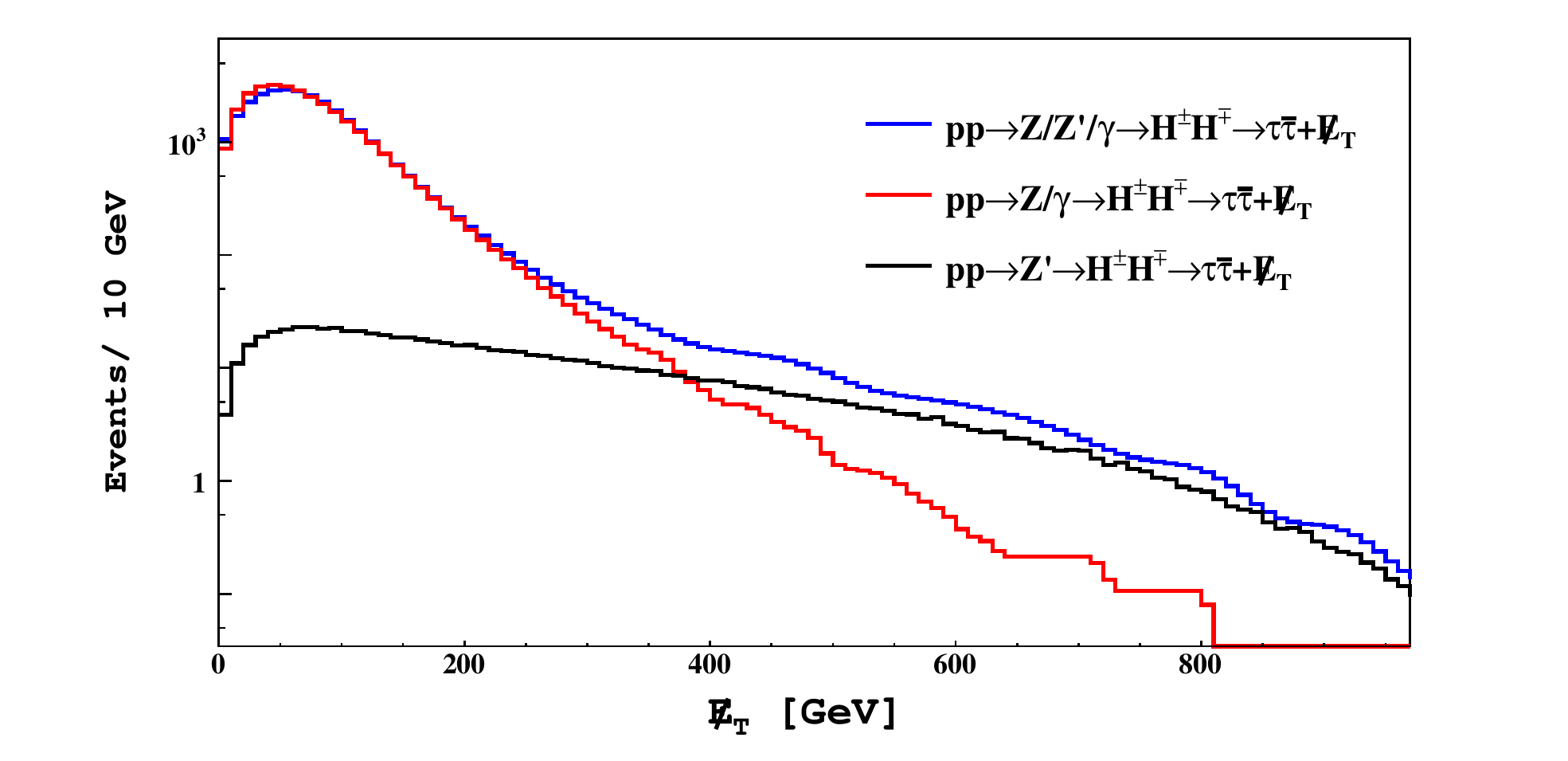} 
\includegraphics[width=0.5\textwidth]{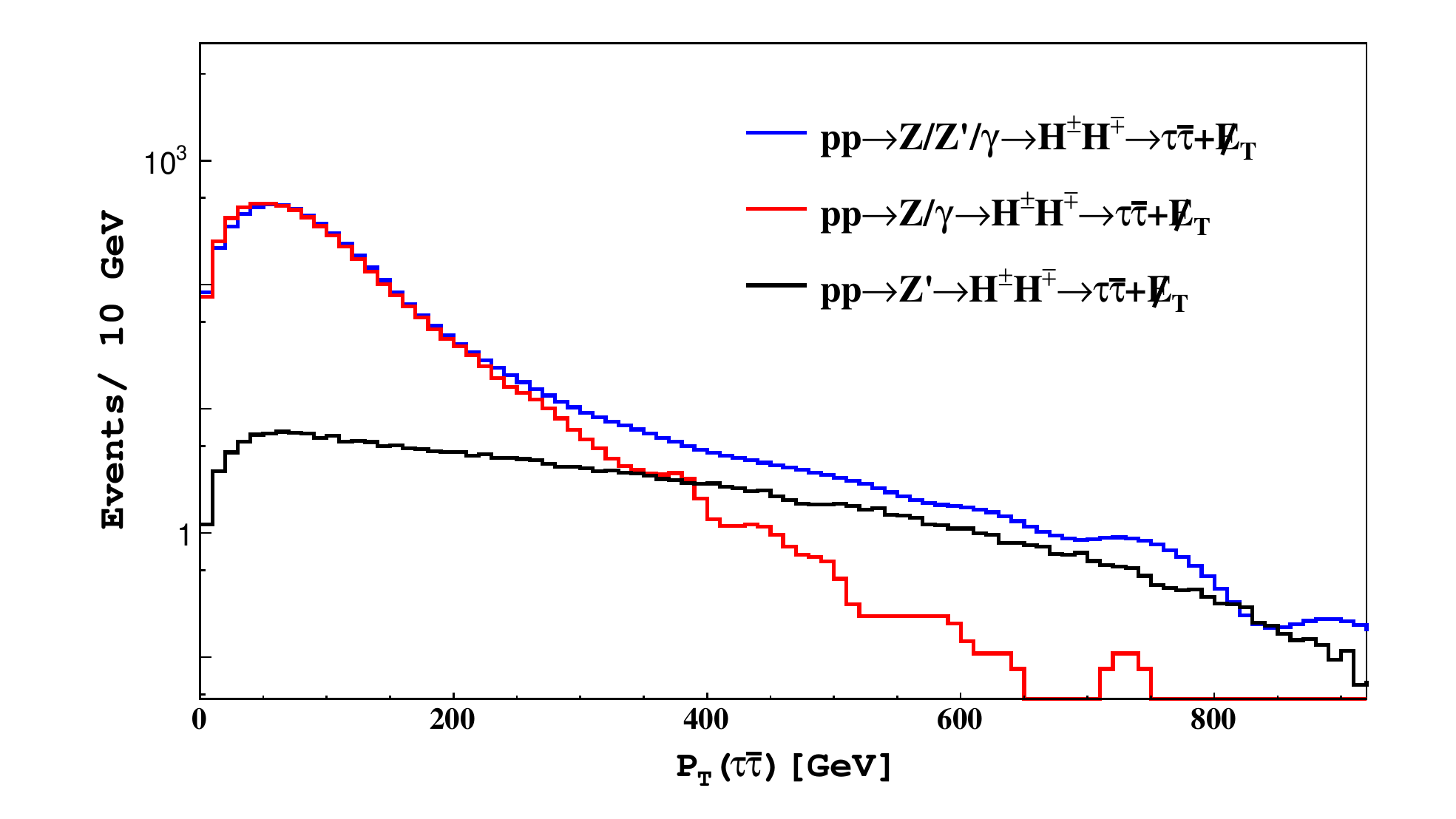}
\vspace{-0.7cm}
\caption{Differential distribution in missing transverse energy (top) and transverse momentum of the $\tau\bar\tau$ pair (bottom)  in the processes 
$pp\to\gamma^*,Z^*,Z' \to H^+ H^-\to \tau\bar\nu_\tau\bar\tau\nu_\tau$,
$pp\to\gamma^*,Z^* \to H^+ H^-\to \tau\bar\nu_\tau\bar\tau\nu_\tau$ and
$pp\to Z' \to H^+ H^-\to \tau\bar\nu_\tau\bar\tau\nu_\tau$.
Here we have used BP3 in Tab.~\ref{table:0} with $\sqrt{s}=14$~TeV and an integrated luminosity of 300~fb$^{-1}$. }
\label{fig:hp1_aa_nocut}
\vspace{-0.7cm}
\end{center}
\end{figure}

This beneficial effect of such an interference is also seen in the differential distributions, e.g., in the $H^+H^-\to \tau\bar\nu_\tau\bar\tau\nu_\tau\to 
\tau\bar\tau+\met$ channel, where $\met$ represents the missing transverse energy due to neutrinos escaping detection. This is illustrated 
in Fig.~\ref{fig:hp1_aa_nocut}.
In the plots, we show the spectra of the total missing transverse energy  (top frame) and  transverse momentum of the $\tau\bar\tau$ system (bottom frame). From here, it is clear that the contribution of a very massive $Z^\prime$, combined with its interference with $\gamma$ and $Z$, has a twofold effect. On the one hand, the total cross section for $pp\to Z^\prime\to H^+ H^-\to {\tau}\bar\tau +\met$ at 14~TeV, which  is already a significant $2.5\times 10^{-3}$~pb, through the effect of the  interference is  enhanced  by more than one order of magnitude. On the other hand, the presence of  the $Z'$  pushes the final state particles to the high end of these distributions, which is not the case for the MSSM wherein the final state particles cannot be extracted from the huge irreducible background existing at low values of these kinematic observables. Hence, by imposing a minimum requirement on $\met$ and/or $p_T({\tau\bar\tau})$ of several hundreds of GeV, one should be able to extract a BLSSM signal, so long that reducible backgrounds are also controlled at the same time (which we will show being the case later on). The drawback of this approach is that event rates for  the signal might turn out be rather small in the end (notice the normalization of the curves in Fig.~\ref{fig:hp1_aa_nocut}), so that event samples generated  by the HL-LHC may indeed be needed to pursue this analysis. Indeed, in this case, the tenfold increase in all
event rates will enable us to probe at the same time not only the fully tauonic signature of charged Higgs bosons, i.e., 
 $pp\to\gamma^*,Z^*,Z' \to H^+ H^-\to \tau\bar\nu_\tau\bar\tau\nu_\tau$, but also the semi-hadronic one, i.e.,  $pp\to\gamma^*,Z^*,Z' \to H^+ H^-\to jj\tau\nu_\tau$. 

\begin{table*}[!t]
\begin{tabular}{|c|c|c|c|c|c|c|c|c|}
\hline
Cuts & BP1 &BP2 &BP3&BP4&BP5& $W^+W^-$ & $t\bar{t}$ & $\gamma^\ast, Z^*$\\
\hline 
$|\eta(j)|<2.4$ & 3325&11820 &27474& 96022 &22356& 130744  & 1594233  & 202381600  \\
\hline
$|\eta(l)|<2.5$ &3276& 11627 & 22789 &93657 & 21940 & 120735  & 1156550  & 197371573    \\
\hline
{{$p_T(j){{>30}}$~GeV}}  &2261 &7908 &15297 &59591 &14582 &66943   & 560267  & 180020149  \\
\hline
{{$p_T(l){{>}}30$~GeV}} & 1608  & 5359&9789.3 &  33190& 9554&24645   & 318578  & 111009979   \\
\hline
 $b$-jet veto & 1591 &5300  & 9694&32945 &9461 &24519   & 161669  & 110349936   \\
\hline
 $\met>350$~GeV &42 (27)&91 (59)&85 (54) &97 (58) &99 (66) &3   & 99 & 0\\
\hline
\end{tabular}
\caption{The cut flow for the full process $pp\to \gamma^\ast,Z^\ast,Z^\prime\to H^+ H^-\to
  \tau\bar\tau+\met$ for our 5 BPs at $\sqrt{s}=14$~TeV and an integrated luminosity of  300~fb$^{-1}$. The last three columns correspond to the relevant reducible backgrounds: the first column for $W^+W^-$, the second column for $t\bar{t}$  and the third column for the Drell-Yan (DY) process. For all 5 BPs, in the last line, the yield of the full process 
is shown alongside that of the signal rate only (in paratheses), as defined in the text. 
} 
 \label{table:1}
\end{table*}

But let us now proceed to the signal-to-background analysis.  Both signal and backgrounds are computed with MadGraph5 \cite{Alwall:2014hca} that is used to estimate multiparton amplitudes and to generate events for the calculation of the cross sections as well as for their 
subsequent processing. PYTHIA6 \cite{Sjostrand:2007gs} has been used for showering, hadronisation, heavy flavour decays and for adding the soft underlying event. The simulation of the response of the ATLAS and CMS detectors was done with the DELPHES package \cite{deFavereau:2013fsa}, wherein  reconstructed objects are simulated from the parametrized detector response and includes tracks, calorimeter deposits and high level objects such as isolated electrons, jets, taus and missing transverse momentum. Finally, for  event reconstruction, we have  used MadAnalysis5 \cite{Conte:2012fm}. 

\begin{figure}[!h]
\begin{center}
\includegraphics[width=0.45\textwidth]{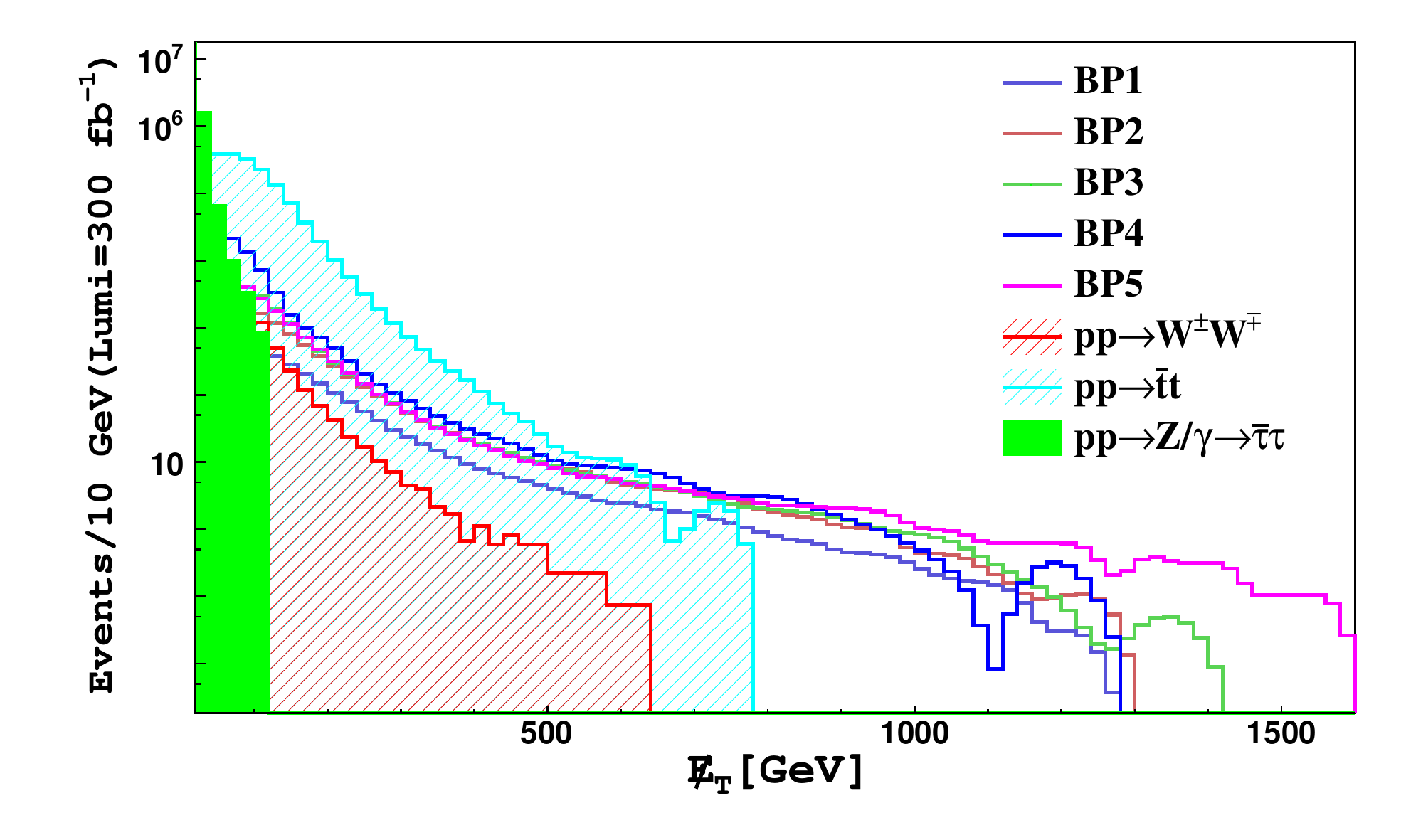}
\includegraphics[width=0.45\textwidth]{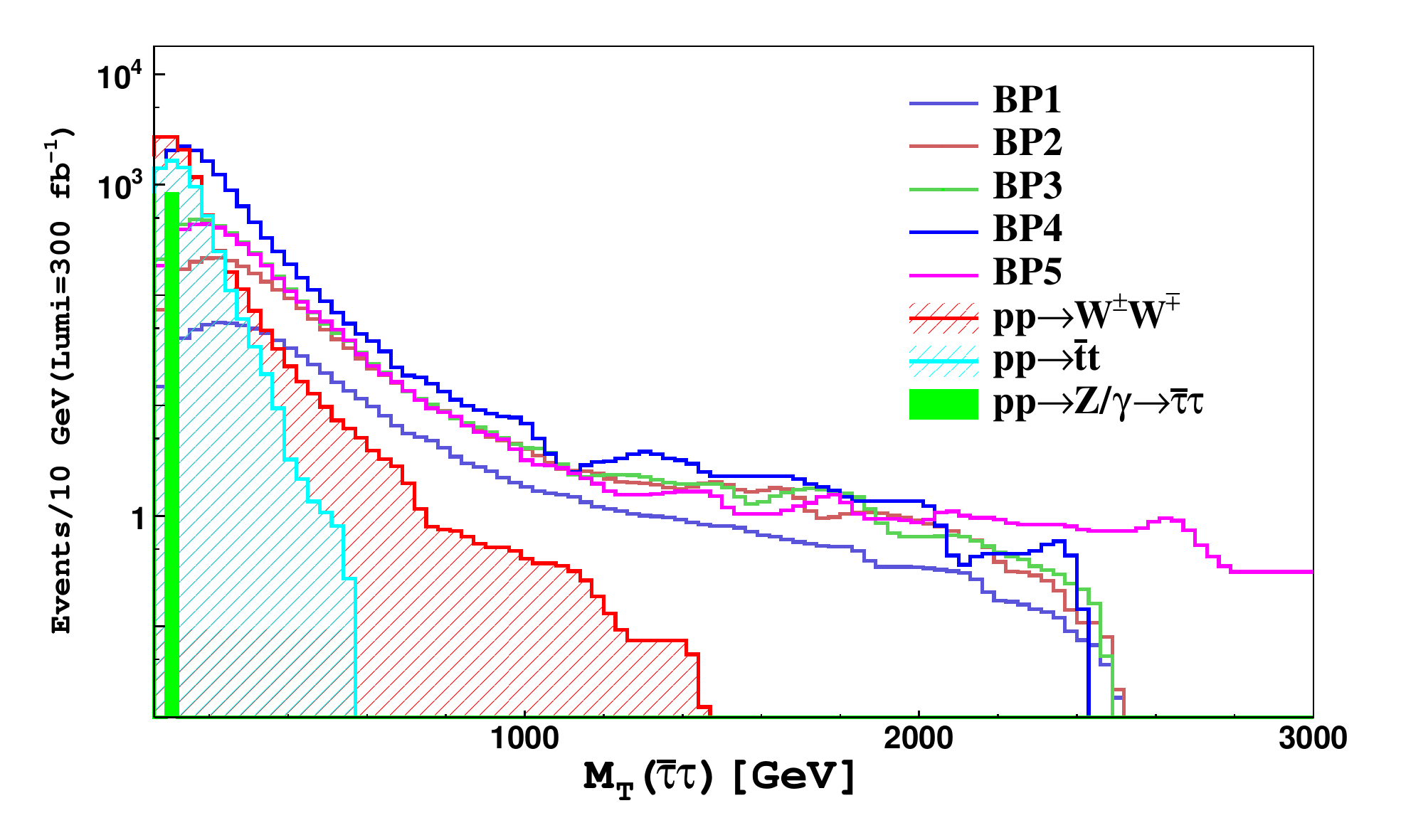}
\vspace{-0.3cm}
\caption{Missing transverse energy distribution for the full process $pp\to \gamma^*,Z^*,Z'\to H^+ H^-\to
  \tau\bar \tau+\met$ and its backgrounds at $\sqrt{s}=14$~TeV and an integrated luminosity of  300~fb$^{-1}$. No cuts are enforced here.}
\label{Fig:6}
\end{center}
\end{figure}

First, we study the fully tauonic decays of the charged Higgs boson pair. As for the $\tau$'s, we use the $\tau$-tagging algorithm included in MadAnalysis5  so that both leptons and jets  are identified as originating from a $\tau$ if they
 can be matched to it when lying within a cone of radius  $\Delta R=0.4$ around a parton-level $\tau${{, as well matching the charged tracks from the $\tau$ decays}}, this yielding an overall efficiency of about $40\%$. Further, the missing transverse
energy $\met$ in the event is defined as the negative sum of the
transverse momentum of all reconstructed objects, so that the quality of 
the reconstruction of all charged particles, especially jets and electrons, has strong bearings on  the unwanted amount of missing energy.
Notice that the presence of two neutrinos associated with
the final state $\tau$'s makes it impossible to reconstruct the $H^\pm$ mass, however, one can instead reconstruct a Jacobian peak which should be correlated to the $Z'$ mass, e.g., through a transverse mass distribution, $M_T$, defined by using all visible objects in the detector and the $\met$. 
Fig.~\ref{Fig:6} shows both the $\met$ and 
$M_T$ observables, prior to any cut, illustrating that they  correlate equally to the actual value of $M_{Z'}$.
The cut flow we have exploited is found in Tab.~\ref{table:1}, wherein $l=e,\mu$ {{and $\tau$}}, $\eta$ and $p_T$ refer to pseudorapidity and transverse momentum, respectively, and the $b$-jet veto is enforced by rejecting events that contain at least one $b$-tagged jet. 
The dominant (reducible) background processes are $t\bar{t}$ with leptonic decays  (which can in particular be reduced
by the aforementioned veto against the existence of a high $p_T$ bottom-quark jet, tagged as such), SM  di-boson production
and the DY channels (all  proceeding via $\tau$'s), while $Z+$jets and $W+$jets can be neglected.  The complete $pp\to \gamma^*,Z^*,Z'\to H^+H^-$ process can be established and, as illustrated in the last line of the table, the pure $S$ component in it is extractable as a clear excess above the intrinsic (irreducible) $B$ yield. A cut in $\met$, based on the top frame of Fig.~\ref{Fig:6}, is crucial to achieve this outcome. Finally, the value of $M_{Z'}$ can be fit to the surviving $M_T$ distribution upon subtracting all backgrounds, including the intrinsic (irreducible) one. 

\begin{figure}[!h]
\begin{center}
\includegraphics[width=0.45\textwidth]{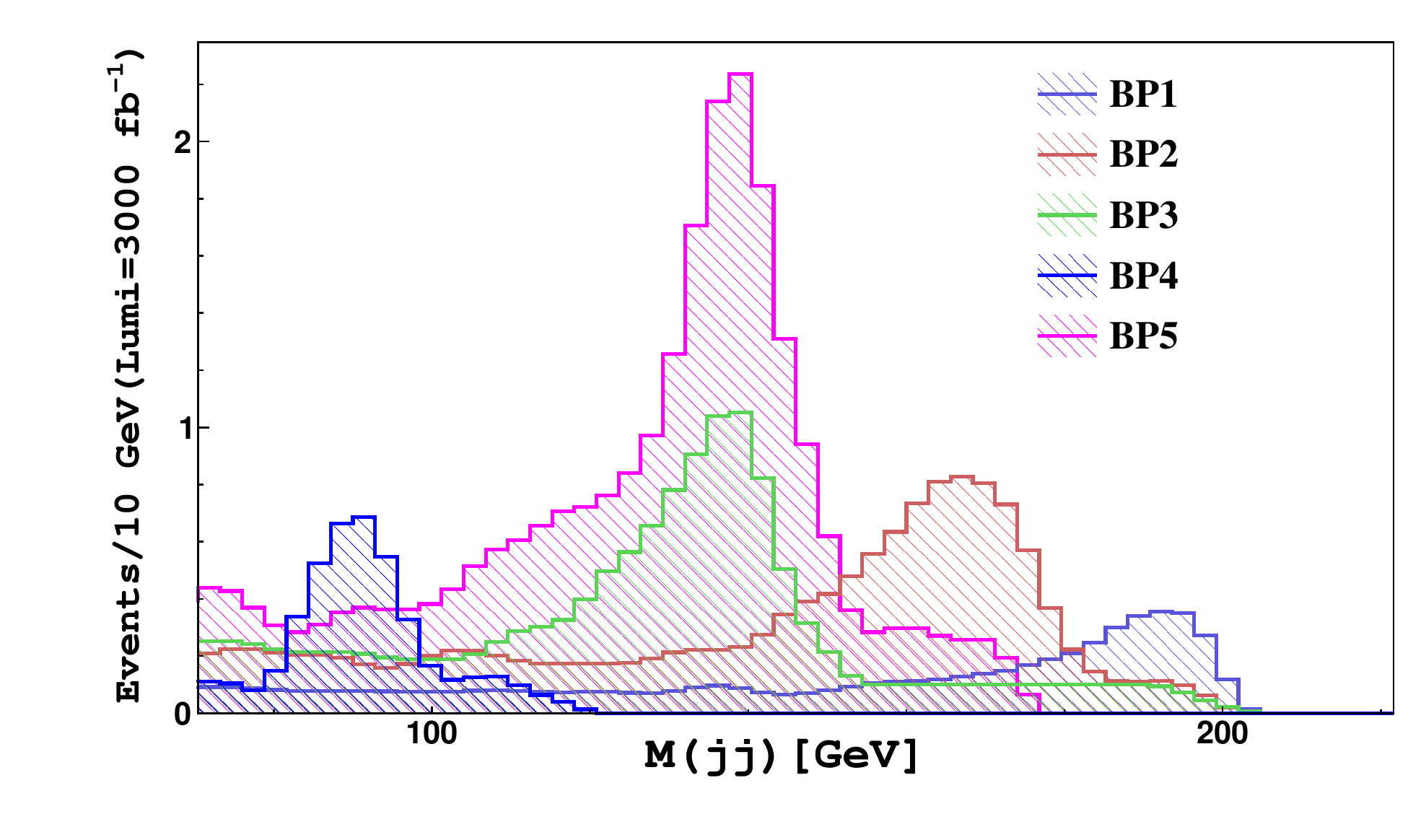}
\vspace{-0.5cm}
\caption{Invariant mass distribution of two jets for the full process $pp\to \gamma^*,Z^*,Z'\to H^+ H^-\to jj\tau+\met$ at $\sqrt{s}=14$~TeV and an integrated luminosity of 3000~fb$^{-1}$. All cuts in Tab.~\ref{table:2} are enforced here. }
\label{fig:Mjj}
\vspace{-0.3cm}
\end{center}
\end{figure}

Then, we probe the signature $\tau \nu_{\tau} j j$ out of full di-charged Higgs boson production and decay $pp\to \gamma^*,Z^*,Z^\prime\to H^+ H^-\to jj\tau\nu_\tau$. The dominant (reducible) SM background arises from events with $W^\pm$ and $Z$ bosons produced in association with jets. Additional sources of SM background  come from di-boson $VV$ and  $t\bar{t}$ production and semileptonic decays via $\tau's$. For signal isolation we have chosen the essential cuts first introduced by Ref.~\cite{Aaboud:2017fgj}, with $|\eta(l)| < 2.4, \ |\eta(j)| < 2.5, \ p_{T}(l)>30$~GeV and $p_{T}(j) > 30$~GeV. Further, other than (a somewhat returned) $\met$ cut, here, also additional cuts on the di-jet invariant mass, $\Delta R$ separation between $\tau$ and jet as well as $\tau$ transverse momentum are necessary to establish the full $pp\to \gamma^*,Z^*,Z'\to H^+H^-$ process, indeed, in a nearly (reducible) background free environment, see Tab.~\ref{table:2}. Again, the last line of the table makes evident  the pure $S$ component above the intrinsic (irreducible) $B$ yield. In this case, one of the two $H^\pm$ masses is reconstructible, from the di-jet system. This is illustrated in Fig.~\ref{fig:Mjj}, where the $M_{jj}$ variable (here plotted after all cuts) is defined  by choosing the two highest transverse momentum jets. The charged Higgs peaks are most evident around the generated $H^\pm$ mass. Their normalization, upon subtraction of the intrinsic (irreducible) background, would represent the BLSSM specific signal, due to $Z'$ mediation, though this will require a very large luminosity, typical of the HL-LHC \cite{Gianotti:2002xx}.


\begin{table*}
\begin{tabular}{ |c|c|c|c|c|c|c|c|c|c| }
 \hline
  Cuts &BP1&BP2&BP3&BP4&BP5&$W^\pm+$jets &$t\bar{t}$& $VV$ & $Z+$jets\\
 \hline
 $|\eta (l)|  < 2.4 $&548.7 &1765.8& 8395.4&13116 & 6984 &972432764 &77399635  &933376206 &229385192 \\
 \hline
$|\eta (j)|< 2.5 $& 547 &1757 & 8155&12789 &6926 & 858600219 &57376760 &821994013 &201066711 \\
 \hline
$p_T (j)> 30$~GeV& 546&1748 &7942 &12053 &6845 &673668930 &37328371  &628744946 & 187767403  \\
 \hline
$p_T (l)> 30$~GeV& 425&1304 &5872&7002 & 4800& 475798328 &15743232    &463566283 &160922594  \\
 \hline
$\met> 300$~GeV&71& 158&152 & 203&354 & 3855562  &19618   &383327&12402 \\
 \hline
$M(jj)< 200$~GeV& 53& 122&106 &166 & 271& 3572907   &13257   &355569&11162 \\
 \hline
$\Delta R(\tau, j)>2$& 6 &16 & 18&22 & 34& 23092  &372  &3304&0 \\
 \hline
$p_T(\tau) > 90$~GeV&3 (2) &8 (5)  & 9 (6) &11 (7) &17 (12) & 1  &2   &2&0 \\
 \hline
\end{tabular}
\caption{The cut flow for the full process  $pp\to \gamma^*,Z^*,Z'\to H^+ H^-\to jj\tau+\met$  for our 5 BPs at $\sqrt{s}=14$~TeV and an integrated luminosity of  300~fb$^{-1}$.   The last four columns correspond to the relevant reducible backgrounds: the first column for $W^\pm+$jets, the second column for $t\bar{t}$, the third column for $VV$ ($V=W^\pm,Z$) and the fourth column for $Z+$jets.  For all 5 BPs, in the last line, the yield of the full process 
is shown alongside that of the signal rate only (in paratheses), as defined in the text. }  
\label{table:2}
\end{table*}  
\vspace{-1cm}
\section*{Conclusions}
The LHC at CERN will enable during Run 2 and 3 to establish a specific BLSSM signal, mediated by on-shell production of a heavy $Z'$ state and yielding a charged Higgs boson pair, eventually decaying into two $\tau$'s and $\met$. Its HL-LHC version, benefiting from a tenfold increase in instantaneous luminosity, will further allow one to access the final state in which one $H^\pm$ decays hadronically into two jets, the other again going into a $\tau\nu_\tau$ pair. As $H^+H^-$ production in the MSSM is only mediated by an off-shell $\gamma^*,Z^*$ current, such a hallmark BLSSM signature clearly requires sampling the $H^\pm$ decay products at large $\met$ values, which is indeed possible via a judicious choice of experimental cuts.  
\vspace{-0.41cm}
\section*{Acknowledgments}
W.A. would like to thank the Department of Atomic Energy (DAE) Neutrino Project of Harish-Chandra Research Institute. The work of A.H. is supported by
the Swiss National Science Foundation. The work of S.K. is partially supported by the STDF project 18448 and the European Union FP7 ITN Invisibles (Marie Curie Actions, PITN-GA-2011-289442). The work of S.K. and S.M. was partially supported by the H2020-MSCA-RISE-2014 grant No. 645722 (NonMinimalHiggs). S.M. is financed in part through the NExT Institute.

\end{document}